\newcommand   {\about} {\mbox{$\sim$}}
\newcommand   {\mic}   {\mbox{$\mu$m}}

\newcommand   {\arcs}  {\mbox{$^{\prime\prime}$}}

\newcommand {\ga}    {\mbox{\rlap{\hbox{\lower4pt\hbox{$\sim$}}}\hbox{$>$}}}
\newcommand {\la}    {\mbox{\rlap{\hbox{\lower4pt\hbox{$\sim$}}}\hbox{$<$}}}

\documentstyle[11pt,paspconf]{article}

\begin{document}

\title{Submillimeter Imaging of NGC 891 with SHARC}
\author{E. Serabyn, D.C. Lis, C.D. Dowell, D.J. Benford, T.R. Hunter}
\affil{Caltech 320-47, Pasadena, CA 91125}

\author{M. Trewhella}
\affil{IPAC, Caltech 100-22, Pasadena, CA 91125}

\author{S.H. Moseley}
\affil{NASA-Goddard Space Flight Center, Greenbelt, MD 20771}

\begin{abstract}
The advent of submillimeter wavelength array cameras operating on
large ground-based telescopes is revolutionizing imaging at these
wavelengths, enabling high-resolution submillimeter surveys of dust
emission in star-forming regions and galaxies. Here we present a
recent 350 $\mu$m image of the edge-on galaxy NGC 891, which was
obtained with the Submillimeter High Angular Resolution Camera (SHARC)
at the Caltech Submillimeter Observatory (CSO). We find that high
resolution submillimeter data is a vital complement to shorter
wavelength satellite data, which enables a reliable
separation of the cold dust component seen at millimeter wavelengths
from the warmer component which dominates the far-infrared (FIR) luminosity.
 
\end{abstract}

\keywords{Brevity,models}

\section{Introduction}

The recently developed submillimeter wavelength camera SHARC (Wang
et al. 1996) has established the viability of long-wave imaging
cameras (to about 500 $\mu$m) 
which are based on the short-wave prescription of illuminating
planar detector arrays with geometrically defined beams. 
Wide-field, high-resolution imaging at 
wavelengths just longward of the FIR regime accessible only
from space-based platforms then becomes possible. 
Combined also with even longer wavelength
submillimeter and millimeter wave data, the accurate delineation of dust
emission spectra in galaxies and star-forming regions then becomes viable.
Specifically, 350 $\mu$m data is vital to the determination of the
properties of cold dust (5-20 K) because this wavelength lies near
the peak emission wavelength for such dust, and so allows 
for a much more accurate assessment of dust parameters than
can be had solely from distant longer or shorter wavelength observations.

\begin{figure}
\includegraphics{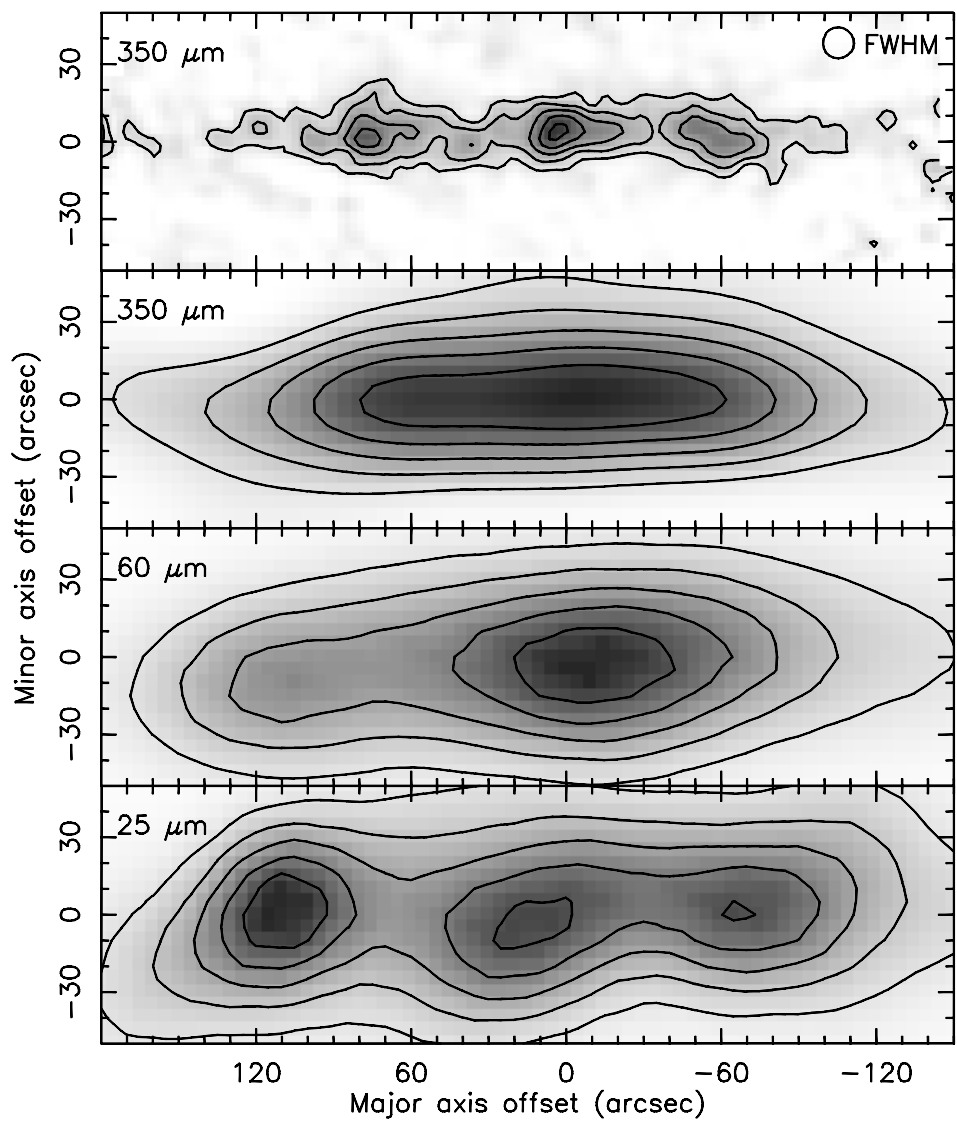}
\vspace{14.5cm}
\caption{Submillimeter and FIR maps of NGC 891. Top panel: SHARC 350
  $\mu$m image of NGC 891, at a resolution of 12$^{\prime\prime}$. The
  peak flux on the nuceus is 5.6 Jy, and the rms noise in the map is
  0.55 Jy. The
  image is rotated such that its major axis is horizontal, and positive
  offsets along the major axis are to the northeast. Second through
  fourth panels: Our 350 $\mu$m data and the IRAS HIRES data
  convolved to the resolution of the IRAS 60 $\mu$m map (64 $\times$ 
  45$^{\prime\prime}$ at a -28$\deg$ position angle). The peak fluxes
  in the 3 lower maps are 419, 201, and 15.7 MJy/sr. The contours are
  all 25,40,55,70, and 85\% of the peak values.}
\end{figure}

Herein we use our 350 $\mu$m SHARC mapping results together
with HIRES processed IRAS maps and published 1.3 mm data to constrain
the properties of the dust emission in the edge-on galaxy NGC 891, an
oft-cited analog to our own Milky Way. Fig. 1 presents our 350 $\mu$m
map, acquired in Jan. 1998. Morphologically it is
similar to the 1300 $\mu$m map of Gu\'elin et al. (1993): 
three peaks are resolved along
the plane of the galaxy, including the nucleus and a peak to either
side attributable to the star-formation ring. The peak flux from the
nucleus is 5.6 $\pm$ 0.55 Jy in a 12$\arcs$ beam, 
and the integrated flux in our
map is 117 Jy. Furthermore, the dust layer is clearly resolved in the
vertical direction, and we derive a FWHM thickness for the dust layer of
900 $\pm$ 100 pc (averaged across the three main peaks).

\begin{figure}
\includegraphics{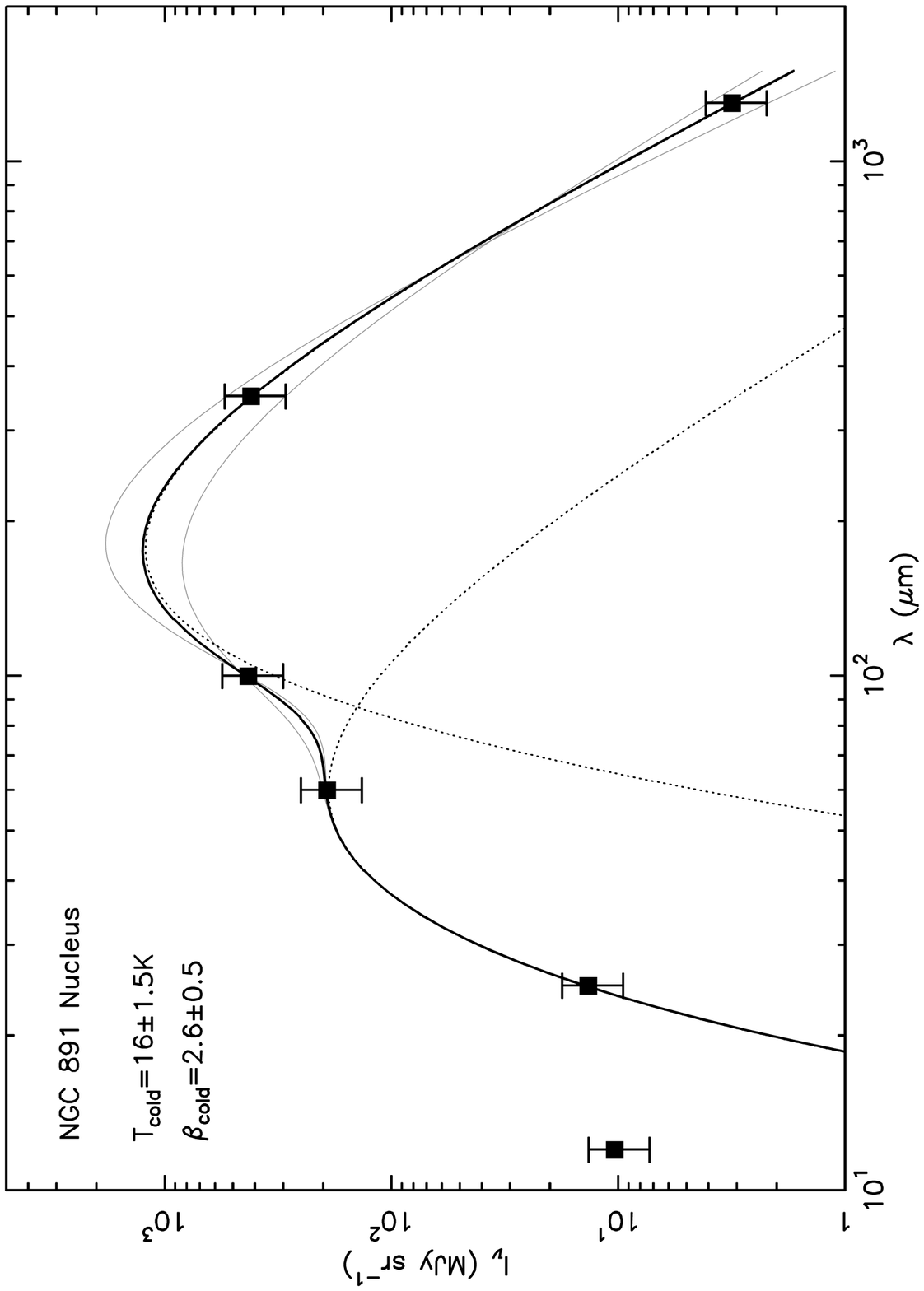}
\vspace{8cm}
\caption{Spectral energy distribution for the nucleus of NGC 891, 
  at a resolution of 45$\arcs$ $\times$ 60$^{\prime\prime}$ (black 
  squares). Together with our 350 $\mu$m point and the IRAS HIRES 
  results, the 1300 $\mu$m point of
  Gu\'elin et al. (1993) is shown. The error bars correspond to an
  absolute calibration uncertainty of 30\%. The solid curve gives the best fit
  two component model, with the two individual components given as 
  dashed curves. Bounding curves to estimate error bars 
  are given in light grey.}
\end{figure}

In order to constrain the temperatures and emissivity law applicable
to the dust in NGC 891, we used HIRES processed IRAS images of the
galaxy together with our 350~\mic\ image and the 1300~\mic\ image. 
The angular resolution of the 60$\mu$m IRAS
image ($\sim 64\arcs \times 45\arcs$ at a --28\deg\ position angle) is
sufficient to resolve several points along the major axis of the
galaxy. All the remaining images were convolved to the angular
resolution of the 60~\mic\ image.  Since the angular resolution of the
100~\mic\ image ($\sim 100\arcs \times 82\arcs$) was too low to be included
directly in the analysis, we assumed a constant 100/60~\mic\ flux
ratio of 0.45 corresponding to the nucleus of the galaxy (this ratio
varies only slightly across the galaxy, between \about 0.4 and 0.45).

Figure~2 shows the observed continuum spectrum
toward the center of NGC891. We have modeled the observed spectrum as a
sum of two modified blackbody functions (e.g. Lis \& Menten
1998). We ignored the 12~\mic\ IRAS data point since the emission
at this wavelength is strongly affected by transiently heated
small dust grains. First we assumed that the 25~\mic\ and 60~\mic\ emission
originates in a `warm' dust component with a grain
emissivity frequency exponent of $\beta = 2$. The dust temperature and the
100~\mic\ optical depth of this component were free parameters in the
fit, along with the temperature, 100~\mic\ optical depth, and $\beta$
for the `cold' dust component, which dominates the emission at $\lambda
\ga 100$~\mic. The
parameters of the cold component are not sensitive to the assumed
value of $\beta$ for the warm component.
The best fit model to the observed spectrum is shown as
a solid black line in Figure~2, along with the spectra of the two
individual components and a pair of limiting curves. 
The best fit parameters for the cold component are $T_d =
16 \pm 2$~K, $\beta = 2.6 \pm 0.5$ (the error bars allow both for
uncertainties at a given location and for variations across
the three peaks), and $\tau_{100} = 2.0$ (to a factor of 2). 
The resultant temperature of the
warm component is \about 50~K (if the 25 $\mu$m point is assumed
to also arise from transiently heated grains, the temperature of the warm
component can be as low as 29 K, and this results in a
cold component temperature lower by a few K) and its optical depth 
(and mass) is a small
fraction ($\sim 3 \times 10^{-4}$) of the total. Similar fits obtained at
several positions along the major axis of the galaxy show that the
parameters describing the cold component show little variation, with
cold component
temperatures varying between 15 and 16~K, the best value of $\beta$
between 2.35 and 2.65, and the 100~\mic\ optical depth varying by
about a factor of 2 for the three peaks. 

To model long-wave galactic dust emission, at least 6 parameters are generally
required: temperatures, long-wave opacity exponents, and reference
frequency opacities for each of 2 temperature components.
Measurement of these parameters has heretofore 
been quite difficult because of
the largely inaccessible region between the longest (100 $\mu$m)
IRAS waveband and the millimeter regime. In particular, for
dust cold enough to show peak emission in the submillimeter, 
these parameters are 
necessarily poorly defined, resulting in the need to simply assign
values to a subset of the 
parameters, such as the temperatures involved (e.g. Gu\'elin et al.
1993). More insidiously, attempts to bridge the submillimeter spectral
gap when a spectral peak is present therein necessarily results in a
significant underestimate of the 
emissivity spectral index, $\beta$, for the coldest component. 
However, the addition of intermediate frequency data closer to the
spectral peak allows
the model parameters to be disentangled to a much greater extent. Indeed, it 
is precisely this
effect which leads to our rather high value for $\beta$: with a well
determined cold-component temperature of only 16 K, the only way to generate 
the large flux increase seen between 1.3 and 0.35 mm is with a
large $\beta$. Thus, the intermediate
frequency data at 350 $\mu$m not only allows for a very
well constrained cold-component temperature, but also for 
a much more reliable emissivity slope estimate. However, the
ramifications also extend beyond the coldest dust component, because 
if a substantial fraction of the 100 $\mu$m flux also arises in such a cold
component, simple 
100/60 $\mu$m color temperatures become very misleading indicators for
the warmer component.


\begin{references}
\reference Wang, N. et al. 1996, Applied Optics, 35, 6629
\reference Gu\'elin, M. et al. 1993, A\&A, 279, L37
\reference Lis, D.C. \& Menten, K.M. 1998, \apj, 507, in press
\end{references}
\end{document}